\documentclass[final]{cimento_arxiv}
\usepackage{graphicx,bm}  

\newcommand{\preprintnumber}{MIT-CTP/5626}

\title{Low-lying baryon resonances from lattice QCD}

\author{
  John Bulava\from{ins:a},
  B\'{a}rbara Cid-Mora\from{ins:b},
  Andrew D. Hanlon\from{ins:c},
  Ben H\"{o}rz\from{ins:d},
  Daniel Mohler\from{ins:e}\from{ins:b},
  Colin Morningstar\from{ins:f}\thanks{Speaker.},
  Joseph Moscoso\from{ins:g},\\
  Amy Nicholson\from{ins:g},
  Fernando Romero-L\'{o}pez\from{ins:h},
  Sarah Skinner\from{ins:f},\\
  Andr\'{e} Walker-Loud\from{ins:i}
 (for the Baryon Scattering (BaSc) Collaboration)}

\instlist{
\inst{ins:a}Fakult\"{a}t f\"{u}r Physik und Astronomie, Institut f\"{u}r Theoretische 
   Physik II, Ruhr-Universit\"{u}t Bochum, 44780 Bochum, Germany
\inst{ins:b}GSI Helmholtz Centre for Heavy Ion Research, Darmstadt, Germany
\inst{ins:c}Physics Department, Brookhaven National Laboratory, Upton, New York 11973, USA
\inst{ins:d}Intel Deutschland GmbH, Dornacher Str. 1, 85622 Feldkirchen, Germany
\inst{ins:e}Institut f\"ur Kernphysik, Technische Universit\"at Darmstadt, 
            Schlossgartenstrasse 2, 64289 Darmstadt, Germany
\inst{ins:f}Dept.~of Physics, Carnegie Mellon University, Pittsburgh, PA 15213, USA
\inst{ins:g}Dept.~of Physics and Astronomy, U.~of North Carolina, Chapel Hill, NC 27516, USA
\inst{ins:h}Center for Theor.~Phys., Massachusetts Inst.~of Technology, Cambridge, MA 02139, USA
\inst{ins:i}Nuclear Science Div., Lawrence Berkeley National Lab., Berkeley, CA 94720, USA}

\shortauthor{J.~Bulava et al.}

\begin{document}

\maketitle

\begin{abstract}
Recent results studying the masses and widths of low-lying baryon resonances in lattice QCD are
presented. The $S$-wave $N\pi$ scattering lengths for both total isospins $I = 1/2$ and $I = 3/2$ are 
inferred from the finite-volume spectrum below the inelastic threshold together with the $I = 3/2$ $P$-wave
containing the $\Delta(1232)$ resonance. A lattice QCD computation employing a combined basis of
three-quark and meson-baryon interpolating operators with definite momentum to determine the
coupled channel $\Sigma\pi$-$N\overline{K}$ scattering amplitude in the $\Lambda(1405)$ region is also 
presented. Our results support the picture of a two-pole structure suggested by theoretical approaches 
based on $SU(3)$ chiral symmetry and unitarity.
\end{abstract}

\section{Overview and Methodology}

Recent results\cite{Bulava:2022vpq,BaryonScatteringBaSc:2023ori} obtained in lattice QCD involving the 
scattering of nucleons and $\Sigma$ baryons with pions and antikaons are presented in this talk.  
The goal of the studies discussed here is to determine properties of some of the low-lying baryon 
resonances, such as the $\Delta(1232)$ and $\Lambda(1405)$.

The properties of hadron resonances are encoded in the spectrum of finite-volume stationary-state 
energies involving the interactions among the decay products.  In lattice QCD, the finite-volume
spectrum of the appropriate symmetry channels is first determined, then the scattering $K$-matrix
is parametrized, and fits to the spectrum through the Luscher quantization condition are
carried out to find best-fit values of the $K$-matrix parameters.  With these in hand,
analytic continuation is used to locate the poles of the transition matrix, which yield the
resonance information.

To determine the finite-volume stationary-state spectrum in lattice QCD, a set of appropriate
interpolating operators $O_i(t)$ must be introduced.  The role of these operators is to create states
with significant overlaps onto the low-lying stationary states through their actions on the QCD vacuum 
$\vert 0\rangle$.  It is important that both single-hadron and two-hadron operators are included.  
Using these operators, a matrix $C_{ij}(t)=\langle 0 \vert O_i(t) O^\dagger_j(0)\vert 0\rangle$ of 
temporal correlations are then evaluated.  With a suitable diagonalization of this correlation matrix, 
the stationary-state energies can be extracted from the exponential fall-offs of the eigenvalues.
The success of such extractions depends crucially on the use of very well designed operators
which produce states with little overlaps onto higher lying states that contaminate the signal.
Much work has been done in the past\cite{Basak:2005aq,Morningstar:2013bda} to design such operators.
Evaluating the correlator matrix elements involving multi-hadron operators requires
techniques to efficiently incorporate time-slice to time-slice quark propagators.
Our computations make use of the stochastic LapH method\cite{Morningstar:2011ka}.

The next step is to parametrize either the $K$-matrix or its inverse, then find best-fit values of 
these parameters by matching the spectrum obtained from the quantization condition
$
  \det(\widetilde{K}^{-1}(E_{\rm cm})-B^{(\bm{P})}(E_{\rm cm}))=0,
$
where $E_{\rm cm}$ is the center-of-mass energy, $\widetilde{K}$ is related to $K$ by threshold 
factors, and $B^{(\bm{P})}$ is
the box matrix for total momentum $\bm{P}$, to the spectrum obtained from lattice QCD.
The above quantization condition and the box matrix elements are discussed in detail
in Ref.~\cite{Morningstar:2017spu}, and references contained therein.

\begin{figure}[t]
\begin{center}
 \includegraphics[width=0.77\textwidth]{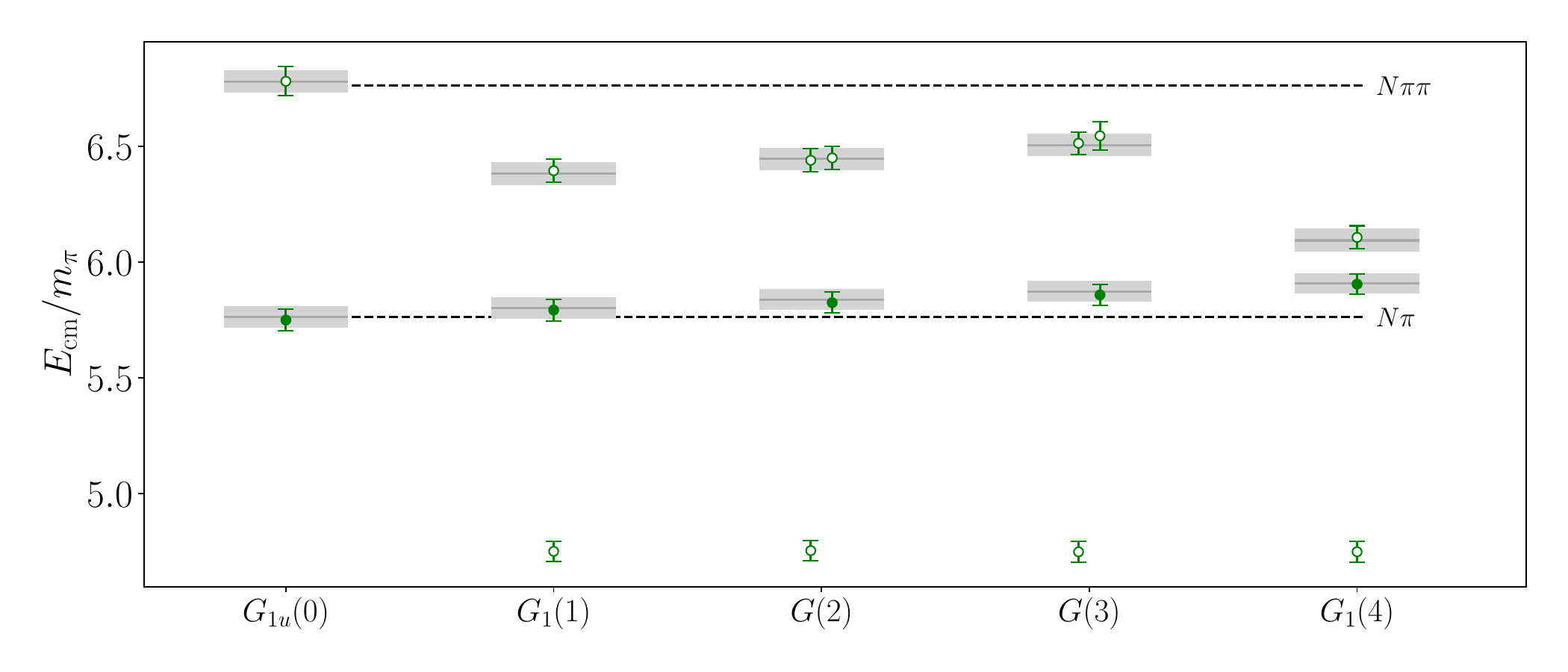}\\
 \includegraphics[width=0.77\textwidth]{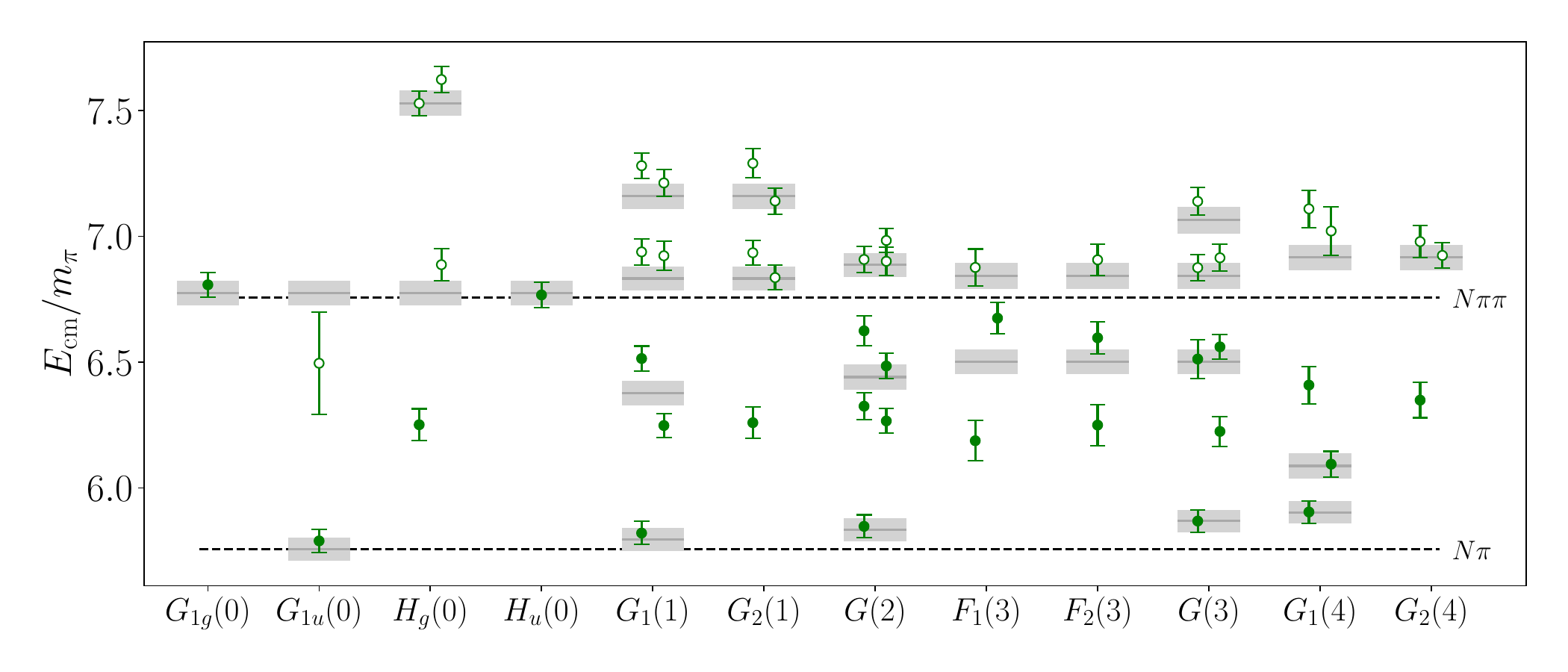}\\
\end{center}
\caption{The low-lying $I=1/2$ (top) and $I=3/2$ (bottom) nucleon-pion spectra in the
center-of-momentum frame on the D200 ensemble. Each column 
corresponds to a particular irrep $\Lambda$ of the little group of total momentum 
$\boldsymbol{P}^2=(2\pi/L)^2\boldsymbol{d}^2$, denoted $\Lambda(\boldsymbol{d}^2)$. Dashed 
lines indicate the boundaries of the elastic region. Solid lines and shaded regions indicate 
non-interacting $N\pi$ levels and their associated statistical errors. Levels employed in 
subsequent fits to constrain the scattering amplitudes are shown with solid symbols. 
Energies are shown as ratios over the pion mass $m_\pi$.
\label{fig:deltaspect}}
\end{figure}

\begin{figure}[t]
\begin{center}
  \includegraphics[width=0.48\textwidth]{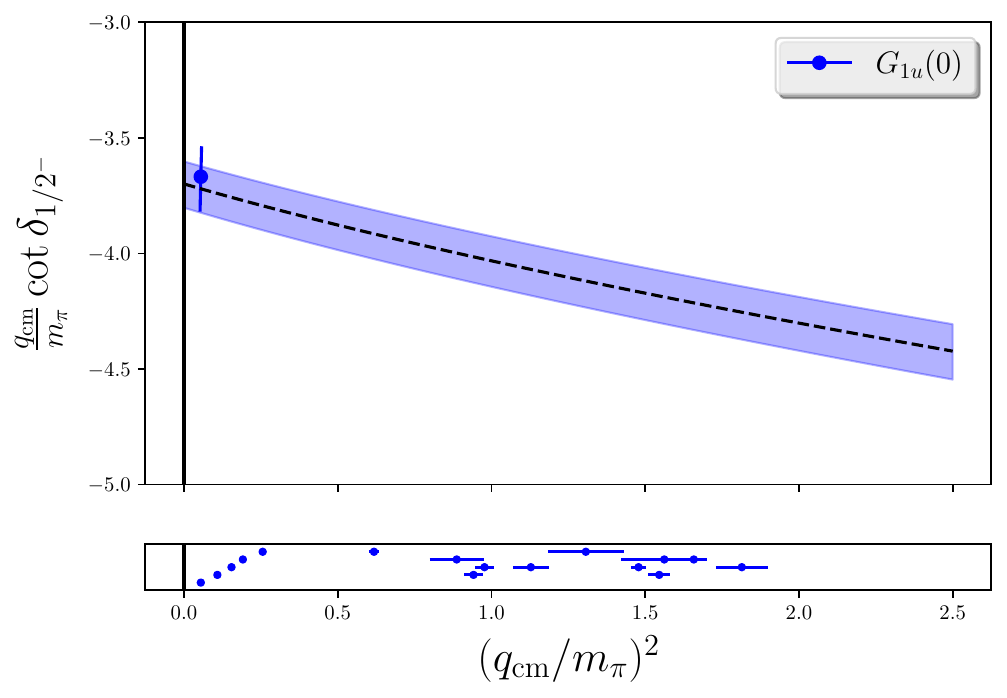}
  \includegraphics[width=0.48\textwidth]{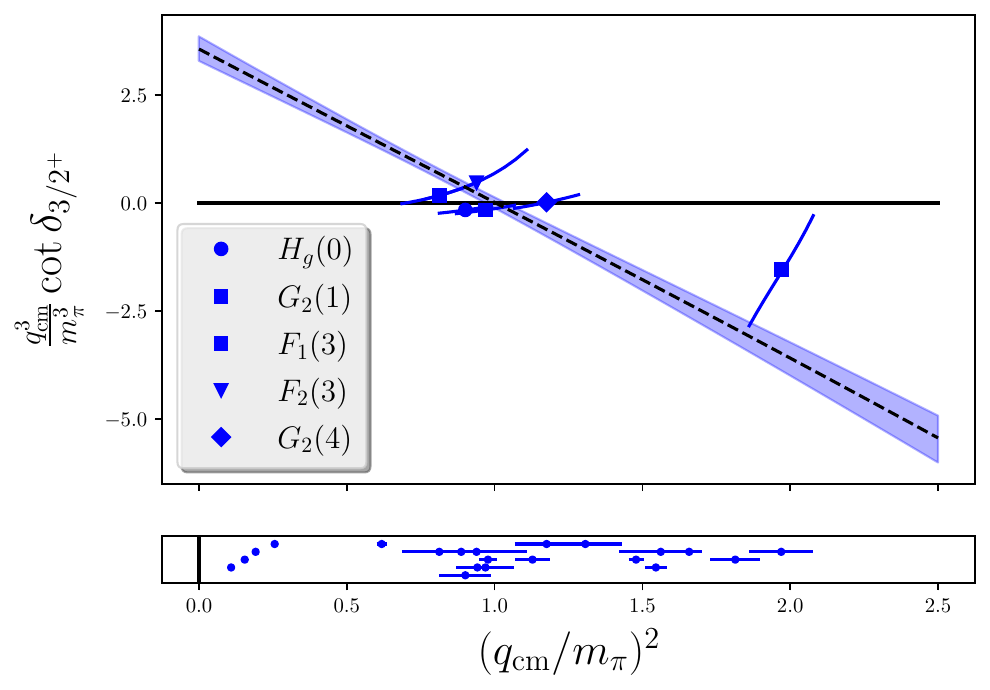}\\
   \includegraphics[width=0.48\textwidth]{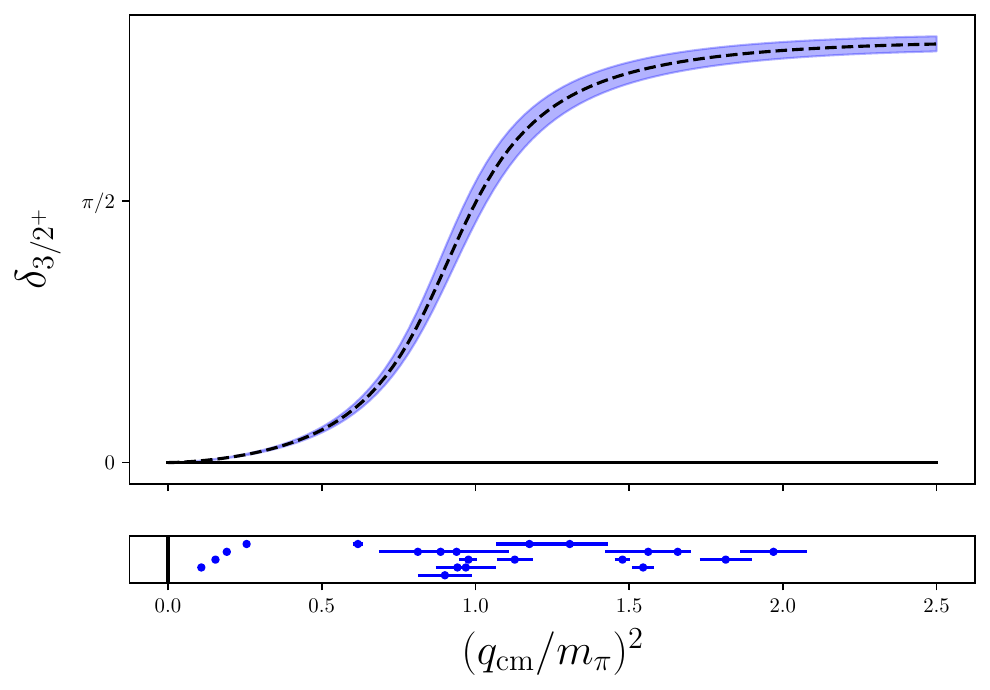}
   \includegraphics[width=0.48\textwidth]{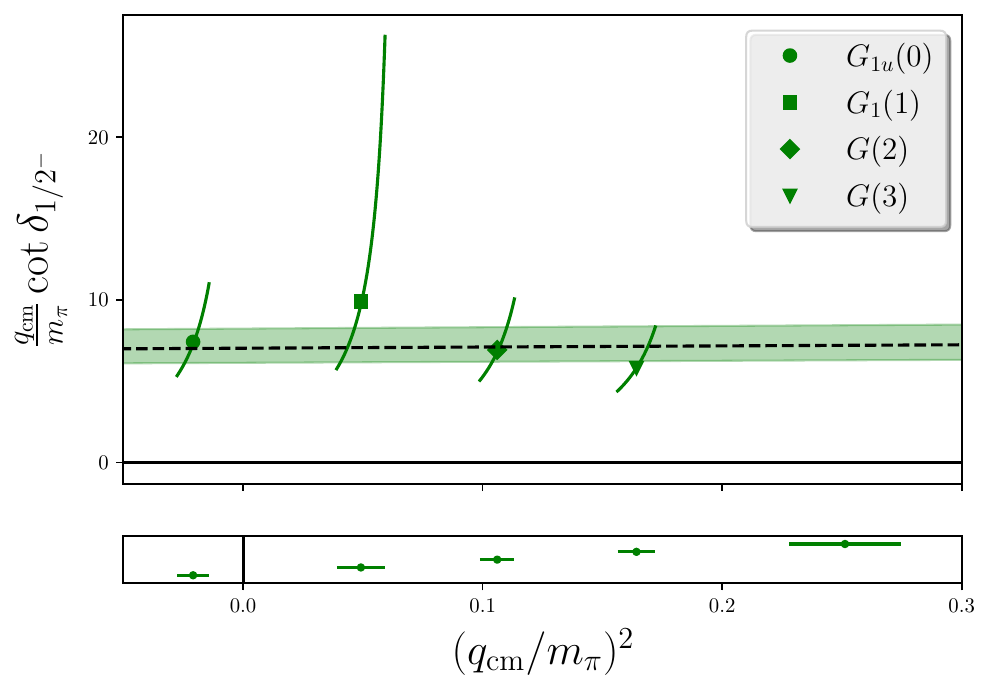}
\end{center}
\caption{The $J^P=1/2^-$ (top left) and $J^P=3/2^+$ (top right) scattering amplitudes
obtained from fits to the $I=3/2$ spectrum in Fig.~\ref{fig:deltaspect}. 
The lower panel of each partial wave shows the squares of the center-of-mass momenta of the
finite-volume levels which contribute to fitting that partial wave.  Most levels, shown with 
solid symbols, contribute to both partial waves, so solving for the partial wave phase shift
shown in the upper panel cannot be done.  When a particular level couples
only to the partial wave shown, a phase shift point can be obtained from the energy level and is shown 
in the upper panel. Hollow symbols indicate such levels. 
(Bottom left) Scattering phase shift of the $I=3/2$, $J^P = 3/2^+$ partial wave containing the $\Delta(1232)$ 
resonance. Levels used in the fit are shown in the lower panel, but no data points are shown in 
the upper panel to more clearly show the final fit form.
(Bottom right) Determination of the scattering length of the $J^P=1/2^-$ wave from
fits to the $I=1/2$ spectrum in Fig.~\ref{fig:deltaspect}.
\label{fig:deltaamps}}
\end{figure}

\begin{figure}[t]
\begin{center}
\includegraphics[width=0.8\linewidth]{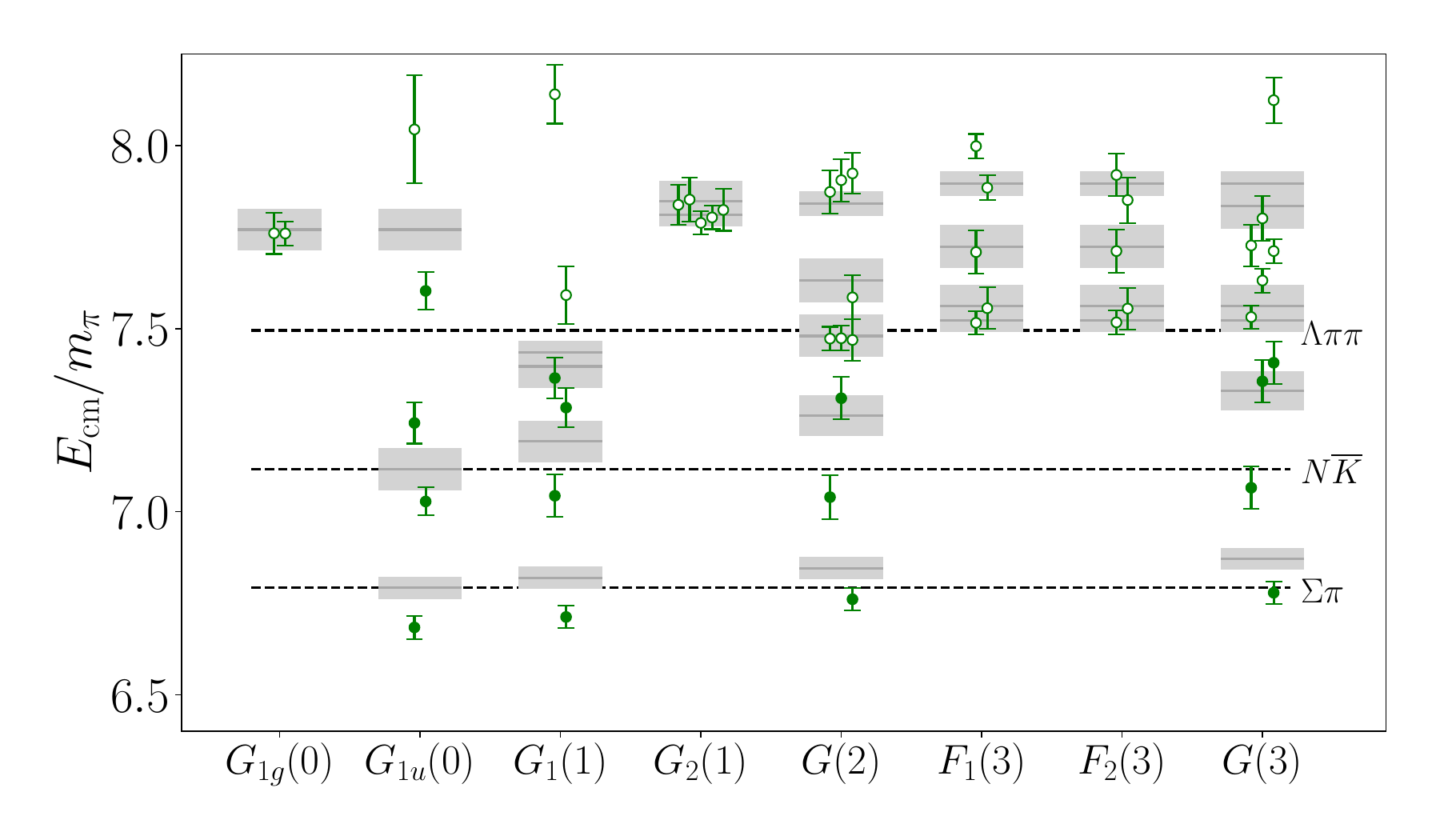}
\includegraphics[width=0.48\linewidth]{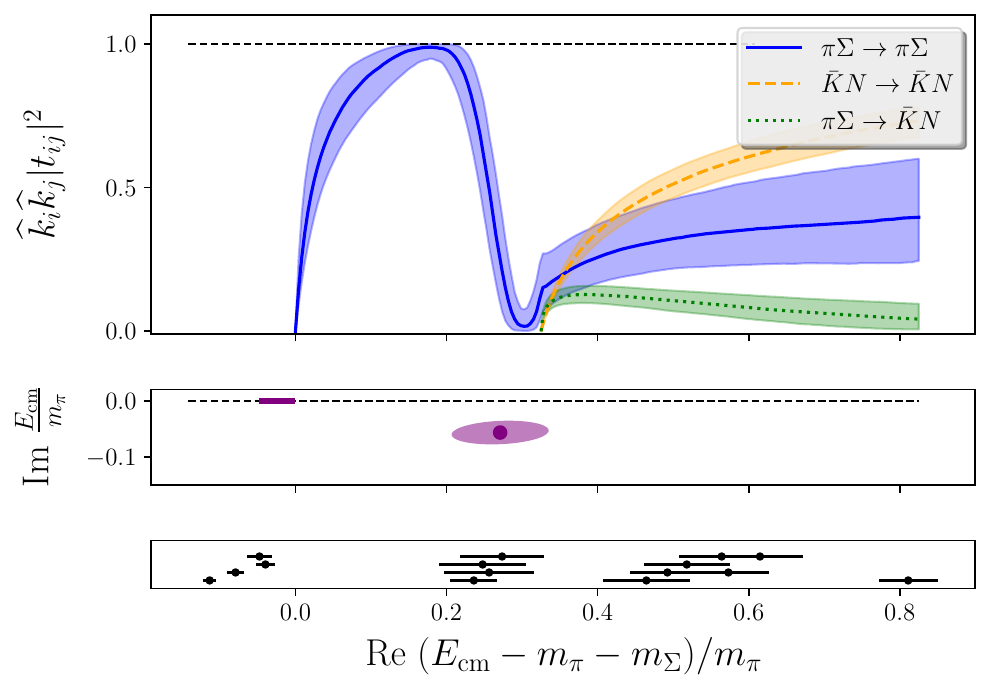}
\includegraphics[width=0.48\linewidth]{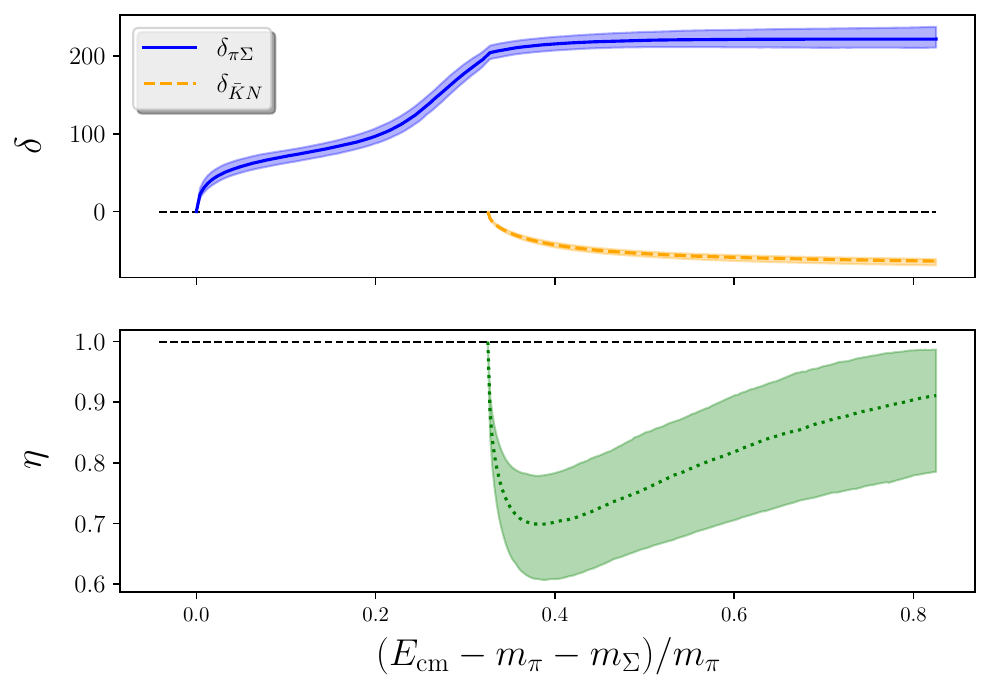}
\end{center}
\caption{(Top) Finite-volume stationary-state energy spectrum, shown as green points, in the 
center-of-mass frame for total isospin $I=0$, strangeness $S=-1$, and various symmetry channels 
indicated along the horizontal axis.  The gray bands show the locations of energy sums
for non-interacting two-particle combinations.  Various two and three particle thresholds
are shown as dashed horizontal lines. 
(Bottom left)
The isospin $I=0$ and strangeness $S=-1$ coupled-channel $\pi\Sigma-\bar{K}N$ transition
 amplitudes as a function of center-of-mass energy difference to the $\pi\Sigma$ threshold.  
  The quantities $t$ and $\hat{k}$ are 
 defined in the text, and the subscripts $i,j$ refer to the flavor channels. 
 The middle panel shows the positions of the $S$-matrix poles in the complex center-of-mass energy 
 plane on the sheets closest to the physical one.  The bottom panel shows the finite-volume 
 spectrum used to constrain the fits involving the transition amplitudes.
(Bottom right)
Inelasticity $\eta$ and phase shifts $\delta_{\pi\Sigma}$ and $\delta_{\bar{K}N}$ 
 against center-of-mass energy difference to the $\pi\Sigma$ threshold.
\label{fig:lambda}}
\end{figure}

\section{Results}

The results shown here use 2000 configurations of the D200 ensemble generated by 
the CLS collaboration.  The lattice is $64^3\times 128$ with lattice spacing 
$a\sim 0.066$~fm, pion mass $m_\pi\sim 200$~MeV, and kaon mass $m_K\sim 480$~MeV.

Results for the spectrum of $N\pi$ states in finite volume are shown in Fig.~\ref{fig:deltaspect}.
The scattering amplitudes obtained using this spectrum are shown in Fig.~\ref{fig:deltaamps}.
The bottom left plot in this figure shows the $\Delta(1232)$ resonance.  The $S$-wave 
isosinglet and isotriplet scattering lengths we obtained are
\begin{equation}
    m_{\pi}a_0^{3/2} = -0.2735(81) \, , \qquad m_{\pi}a_0^{1/2} = 0.142(22),
\end{equation}
and the $\Delta$-resonance mass and width parameters were found to be
\begin{equation}
    \frac{m_{\Delta}}{m_{\pi}} = 6.257(35) , \qquad  g_{\Delta,\rm BW} = 14.41(53),
\end{equation}
where the Breit-Wigner parameter is given by $g_{\Delta,\rm BW}^2 q_{\rm cm}^3\cot(\delta_{3/2^+})
  = 6\pi E_{\rm cm}(m_\Delta^2-E_{\rm cm}^2)$.

The finite-volume spectrum and resulting coupled-channel $\pi\Sigma-\bar{K}N$ transition amplitudes 
for the isospin $I=0$ and strangeness $S=-1$ are shown in Fig.~\ref{fig:lambda}.
The transition elements $t_{ij}^{(J^P)}(E_{\rm cm})$ are defined by
$
t^{-1} = \widetilde{K}^{-1} -i \widehat k, \quad 
$
where $\widehat k = {\rm diag }(k_{\pi \Sigma}, k_{\bar{K}N})$, with 
$k_{\pi \Sigma}, k_{\bar{K}N}$ defined in Ref.~\cite{BaryonScatteringBaSc:2023ori}.
Another way of presenting the results for the amplitudes is to show the scattering phase 
shifts $\delta_i$ and the inelasticity $\eta$, which are shown in the lower right
plot of this figure.  These amplitudes, continued to the complex energy plane, exhibit
a virtual bound state below the $\Sigma\pi$ threshold and a resonance pole just
below the $N\overline{K}$ threshold.  These findings are broadly consistent with
predictions from chiral effective field theory.  The energies of these poles are
\begin{eqnarray}
E_1 &=& 1395(9)_{\rm stat} (2)_{\rm model} (16)_a {\rm  MeV}, \\
E_2 &=&  1456(14)_{\rm stat}(2)_{\rm model}(16)_a - i 
\times 11.7(4.3)_{\rm stat}(4)_{\rm model}(0.1)_a {\rm  MeV}.
\end{eqnarray}
The first errors are statistical, the second are from model variations, and the
last from scale setting.
This is a first-time calculation in lattice QCD of scattering amplitudes in a 
coupled-channel meson-baryon system.

\acknowledgments
Computations were carried out on Frontera~\cite{frontera} at the Texas Advanced Computing 
Center, and at the National Energy Research Scientific Computing Center
using awards NP-ERCAP0005287, NP-ERCAP0010836 and NP-ERCAP0015497.
This work was supported in part by:
the U.S. National Science Foundation under awards PHY-1913158 and PHY-2209167 (CJM and SK), 
PHY-2047185 (AN), and DGE-2040435 (JM);
the U.S. Department of Energy, under grants 
DE-SC0011090 and DE-SC0021006 (FRL), DE-SC0012704 (ADH), DE-AC02-05CH11231 
(AWL); the Mauricio and Carlota Botton Fellowship (FRL);
and the Heisenberg Programme of the Deutsche Forschungsgemeinschaft
project number 454605793 (DM).
NumPy~\cite{harris2020array}, matplotlib~\cite{Hunter:2007}, and the CHROMA software 
suite~\cite{Edwards:2004sx} were used.


\end{document}